# Plasmon modes in bilayer – monolayer graphene heterostructures

Nguyen Van Men[1,2] and Nguyen Quoc Khanh[2]
[1]*University of An Giang, 18-Ung Van Khiem Street, Long Xuyen, An Giang, Viet Nam*
[2]*University of Science - VNUHCM, 227-Nguyen Van Cu Street, 5th District, Ho Chi Minh City, Viet Nam*

______________________________________________________________________

**Abstract**

We investigate the dispersion relation and damping of plasmon modes in a bilayer-monolayer graphene heterostructure with carrier densities $n_{BLG}$ and $n_{MLG}$ at zero temperature within the random-phase-approximation taking into account the nonhomogeneity of the dielectric background of the system. We derive analytical expressions for plasmon frequencies by using long wavelength expansion of response and bare Coulomb interaction functions. We show that optical plasmon dispersion curve of the bilayer-monolayer system lies slightly below that of double-layer graphene (DLG) and the acoustic one lies much lower than that of DLG. We find that while decay rates of acoustic modes of the system and DLG are remarkably different, those of optical modes in both double-layer systems are similar. Except the damping rate of acoustic mode, properties of plasmon excitations in considered system depend remarkably on the interlayer distance, inhomogeneity of the background, density ratio $n_{MLG}/n_{BLG}$ and spacer dielectric constant, especially at large wave-vectors.

______________________________________________________________________

## 1. Introduction

Graphene, a two dimensional material made of sp[2] bonded carbon atoms, attracts a lot of attention in recent years for its excellent electronic properties [1] and possible technological applications [2-5]. Charge carriers in graphene are Dirac-like massless, chiral fermions near the Dirac points, *K* and *K'* where the band structure of graphene is linear. Contrary to monolayer graphene (MLG), bilayer graphene (BLG) has a low-energy parabolic spectrum, although, the chiral form of the effective 2-band Hamiltonian persists [6]. The collective plasmon excitations in the electron gas have been extensively studied and have been used to create plasmonic devices [7]. Due to chiral character of graphene carriers, the plasmons in MLG and BLG exhibit significantly different from those of ordinary two-dimensional electron gas (2DEG) [8-9]. Plasmon dispersion in double-layer systems was revealed to be considerably different from the single layer ones [10-17]. It was shown recently that plasmon modes in MLG-2DEG [18-19] and BLG-2DEG [20] heterostructures even have more interesting properties due to linear electron energy dispersion and chiral character of graphene carriers. Ground state [21] and Coulomb drag [22] in a BLG-MLG heterostructure, which is chiral system, have been investigated recently. To our knowledge, up to now, no calculations on plasmon excitations have been done for BLG-MLG systems, although collective excitations in such massive-massless double-layer structures may have interesting properties. Therefore, in this paper, we consider a double-layer structure consisting of bilayer and monolayer graphene layers isolated from each other by a spacer. Using the random–phase–approximation (RPA) dielectric function, we calculate the plasmon frequency and damping rate in a BLG-MLG heterostructure at zero temperature, taking into account the nonhomogeneity of the dielectric background of the system.

## 2. Theory

We propose a hybrid double-layer system, consisting of BLG and MLG layers immersed in a three-layered dielectric medium with the background dielectric constants $\kappa_1$, $\kappa_2$ and $\kappa_3$, as shown in Fig. 1.

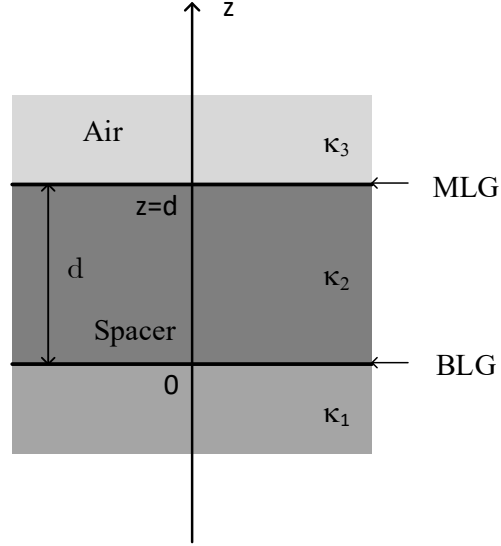

Fig. 1. A BLG-MLG double-layer system embedded in an inhomogeneous dielectric environment with dielectric constants $\kappa_1$, $\kappa_2$ and $\kappa_3$.

The dispersion relations for the longitudinal collective excitations are given by the zeroes of dynamical dielectric function [8-17]

$$\varepsilon(q, \omega_p - i\gamma) = 0 \qquad (1)$$

where $\omega_p$ is the plasmon frequency at a given wave-vector $q$ and $\gamma$ is the damping rate of plasma oscillations. In case of weak damping ($\gamma \ll \omega_p$), the plasmon dispersion and decay rate are determined from the following equations [13, 20]

$$\mathrm{Re}\,\varepsilon(q, \omega_p) = 0 \qquad (2)$$

and

$$\gamma = \mathrm{Im}\,\varepsilon(q, \omega_p) \left( \frac{\partial \mathrm{Re}\,\varepsilon(q,\omega)}{\partial \omega} \bigg|_{\omega=\omega_p} \right)^{-1} \qquad (3)$$

Within RPA, the dynamical dielectric function of a BLG-MLG heterostructure is given by [15-16]

$$\varepsilon_{\text{BLG-MLG}}(q,\omega) = \varepsilon_{\text{BLG}}(q,\omega)\varepsilon_{\text{MLG}}(q,\omega) - v_{12}^2(q)\Pi_{\text{BLG}}(q,\omega)\Pi_{\text{MLG}}(q,\omega) \qquad (4)$$

where $\Pi_{\text{BLG}}(q,\omega)$ ($\Pi_{\text{MLG}}(q,\omega)$) is the zero-temperature response function of BLG (MLG) given in [23] ([8]) and $\varepsilon_{\text{BLG}}(q,\omega)$ ($\varepsilon_{\text{MLG}}(q,\omega)$) is the dielectric function of BLG (MLG),

$$\varepsilon_{\text{BLG}}(q,\omega) = 1 + v_{11}(q)\Pi_{\text{BLG}}(q,\omega) \qquad (5)$$

$$\varepsilon_{\text{MLG}}(q,\omega) = 1 + v_{22}(q)\Pi_{\text{MLG}}(q,\omega) \qquad (6)$$

Here $v_{ii}(q)$ and $v_{ij}(q)$ are respectively the intralayer and interlayer Coulomb matrix elements [16, 22],

$$v_{ij}(q) = \frac{2\pi e^2}{\bar{\kappa}_{ij}(q)q} \qquad (7)$$

where

$$\frac{1}{\bar{\kappa}_{11}} = \frac{2(\kappa_2 \cosh(qd) + \kappa_3 \sinh(qd))}{\kappa_2(\kappa_1 + \kappa_3)\cosh(qd) + (\kappa_1\kappa_3 + \kappa_2^2)\sinh(qd)} \quad (8)$$

$$\frac{1}{\bar{\kappa}_{22}} = \frac{2(\kappa_2 \cosh(qd) + \kappa_1 \sinh(qd))}{\kappa_2(\kappa_1 + \kappa_3)\cosh(qd) + (\kappa_1\kappa_3 + \kappa_2^2)\sinh(qd)} \quad (9)$$

$$\frac{1}{\bar{\kappa}_{12}} = \frac{2\kappa_2}{\kappa_2(\kappa_1 + \kappa_3)\cosh(qd) + (\kappa_1\kappa_3 + \kappa_2^2)\sinh(qd)} \quad (10)$$

Using long wavelength expansions of $\Pi_{MLG}(q,\omega)$ [8] and $\Pi_{BLG}(q,\omega)$ [23] we obtain the following solutions of Eq. 2,

$$\omega_\pm^2 = \frac{1}{2}\left\{ \frac{\omega_{0BLG}^2}{\bar{\kappa}_{11}(qd)} + \frac{\omega_{0MLG}^2}{\bar{\kappa}_{22}(qd)} \pm \sqrt{\left[\frac{\omega_{0BLG}^2}{\bar{\kappa}_{11}(qd)}\right]^2 + \left[\frac{\omega_{0MLG}^2}{\bar{\kappa}_{22}(qd)}\right]^2 + 2\omega_{0BLG}^2\omega_{0MLG}^2 \left[\frac{2}{[\bar{\kappa}_{12}(qd)]^2} - \frac{1}{\bar{\kappa}_{11}(qd)\bar{\kappa}_{22}(qd)}\right]} \right\} \quad (11)$$

where $\omega_{0BLG}^2 = \frac{2\pi e^2 n_{BLG}}{m^*} q$ and $\omega_{0MLG}^2 = \frac{2 v_F e^2 \sqrt{\pi n_{MLG}}}{\hbar} q$ with $v_F$ as the graphene velocity. Here $n_{BLG}$ and $n_{MLG}$ are the carrier density in BLG and MLG, respectively. In limit $q \to 0$ and $qd \ll 1$ we can expand $\frac{1}{\bar{\kappa}_{11}(qd)}$, $\frac{1}{\bar{\kappa}_{22}(qd)}$, and $\frac{1}{\bar{\kappa}_{12}(qd)}$ to the first order $qd$ and have

$$\omega_+^2 = \frac{2(\omega_{0BLG}^2 + \omega_{0MLG}^2)}{\kappa_1 + \kappa_3} \quad (12)$$

and

$$\omega_-^2 = \frac{2qd\,\omega_{0BLG}^2 \omega_{0MLG}^2}{\kappa_2(\omega_{0BLG}^2 + \omega_{0MLG}^2)} \quad (13)$$

The solution $\omega_+ \sim \sqrt{q}$ ($\omega_- \sim q$) corresponds to the optical (acoustic) plasmon mode. Eqs. (12-13) indicate that $\omega_+(q \to 0)$ depends on both $\kappa_1$ and $\kappa_3$, while $\omega_-(q \to 0)$ depends only on $\kappa_2$ as in the case of MLG-2DEG [18] and BLG-2DEG [20] heterostructures. For homogeneous systems ($\kappa_1 = \kappa_2 = \kappa_3 = \kappa$), Eqs. (12) and (13) reduce to

$$\omega_+^2 = \frac{\omega_{0BLG}^2 + \omega_{0MLG}^2}{\kappa} = q\left(\frac{2\pi e^2 n_{BLG}}{\kappa m^*} + \frac{2 v_F e^2 \sqrt{\pi n_{MLG}}}{\kappa \hbar}\right) \quad (14)$$

and

$$\omega_-^2 = \frac{2qd\,\omega_{0BLG}^2 \omega_{0MLG}^2}{\kappa(\omega_{0BLG}^2 + \omega_{0MLG}^2)} = \frac{4\pi e^2 d \frac{n_{BLG}}{m^*} \cdot \frac{v_F \sqrt{\pi n_{MLG}}}{\hbar}}{\left(\frac{\pi n_{BLG}}{m^*} + \frac{v_F \sqrt{\pi n_{MLG}}}{\hbar}\right)\kappa} q^2 \quad (15)$$

We observe that Eqs. (14) and (15) are similar to those given in previous papers [18, 20, 24]. The first (second) term on the r. h. s. of Eq. (14) is the square of the frequency of the optical plasmon oscillations in BLG (MLG).

## 3. Numerical results

We have performed numerical calculations of the frequency and damping rate of plasmon oscillations in a BLG-MLG double-layer system with $\kappa_1 = \kappa_{SiO_2} = 3.8$, $\kappa_3 = \kappa_{air} = 1.0$ and $m^* = 0.033 m_0$ where $m_0$ is the vacuum mass of the electron at zero temperatures, for several values of interlayer separation $d$, dielectric constant $\kappa_2$, $n_{MLG}$ and $n_{BLG}$. We have obtained two solutions of Eq. (2). The higher-frequency (lower-frequency) solution corresponds to in-phase (out of phase) oscillations of densities in the two layers and has been called "optical plasmon" ("acoustic plasmon"). In the following we denote the Fermi energy and Fermi wave-vector of BLG by $E_F$ and $k_F$, respectively.

In Fig. 2(a) we plot the plasmon dispersion for $n_{BLG} = n_{MLG} = 10^{12}\,\text{cm}^{-2}$, $d = 20\,\text{nm}$ and $\kappa_2 = \kappa_{Al_2O_3} = 9.1$. The solid (dashed) lines show the optical (acoustic) plasmon dispersion and the dashed-dotted lines denote the single particle excitation (SPE) boundaries of BLG and MLG. As seen from the figure that the acoustic plasmon dispersion of BLG-MLG system hits the edge of the MLG continuum at a critical wave-vector while the optical one enters MLG SPE region at smaller wave-vector $q \approx 0.24 k_F$. Analytical results given in Eq. (11), displayed by thin lines in Fig. 2(a), indicate that the analytical results are almost identical to the numerical ones at sufficiently small $q$. Fig. 2(b) illustrates the damping rate of plasmon modes in a BLG-MLG heterostructure with $Al_2O_3$ spacer for $d = 20\,\text{nm}$ and $n_{BLG} = n_{MLG} = 10^{12}\,\text{cm}^{-2}$. The figure shows that the damping rate of optical branch increases from zero at $q \approx 0.24 k_F$ where the plasmon dispersion touches the edge of the BLG continuum, gets a peak then decreases to zero and strongly increases when the plasmon enters the continuum of MLG SPE. Fig. 2(b) also indicates that the acoustic mode is totally undamped as the plasmon dispersion lies outside the SPE regions and becomes strongly damped once the plasmon goes into the MLG continuum. Note that in DLG the plasmon dispersion never goes into intraband SPE region and the damping rate of acoustic plasmons decreases to zero when the plasmon dispersion reaches the intraband SPE boundary.

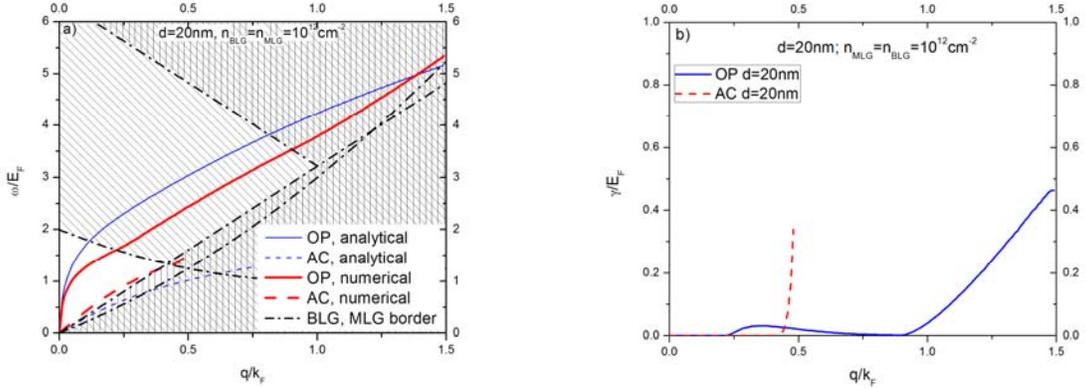

Fig. 2. (a) Plasmon dispersion and (b) damping rate of plasma oscillations in a BLG-MLG system with $\kappa_2 = \kappa_{Al_2O_3} = 9.1$ and $d = 20\,\text{nm}$. The thin solid (dashed) lines are analytical optical (acoustic) results. The dashed-dotted lines show the SPE boundaries of BLG and MLG (color online).

Fig. 3 shows a comparison of plasmon dispersion and damping rate in BLG-MLG, MLG and BLG systems, and in BLG-MLG and DLG double layers for $\kappa_2 = \kappa_{Al_2O_3} = 9.1$, $d = 20\,\text{nm}$ and $n_{BLG} = n_{MLG} = 10^{12}\,\text{cm}^{-2}$. Average background dielectric constant $\bar{\kappa} = \frac{1}{2}(\kappa_1 + \kappa_3) = 2.4$ is used for BLG and DLG. As can be seen from Fig. 3(a) that the plasmon dispersion curve of optical branch of BLG-MLG system lies below (above) that of MLG (BLG). Fig. 3(b) indicates that the plasmon damping rate of optical plasmon mode in BLG-MLG heterostructure is much smaller than that in BLG for the region of intermediate wave-vectors. We observe from Fig. 3(c) that while plasmon dispersion curve of optical branch lies slightly below that of DLG, the acoustic one lies much lower than that of DLG. From Fig. 3(d) we find that the decay rates of optical plasmon modes in BLG-MLG and DLG double layers are quite similar while those of acoustic ones are remarkably different.

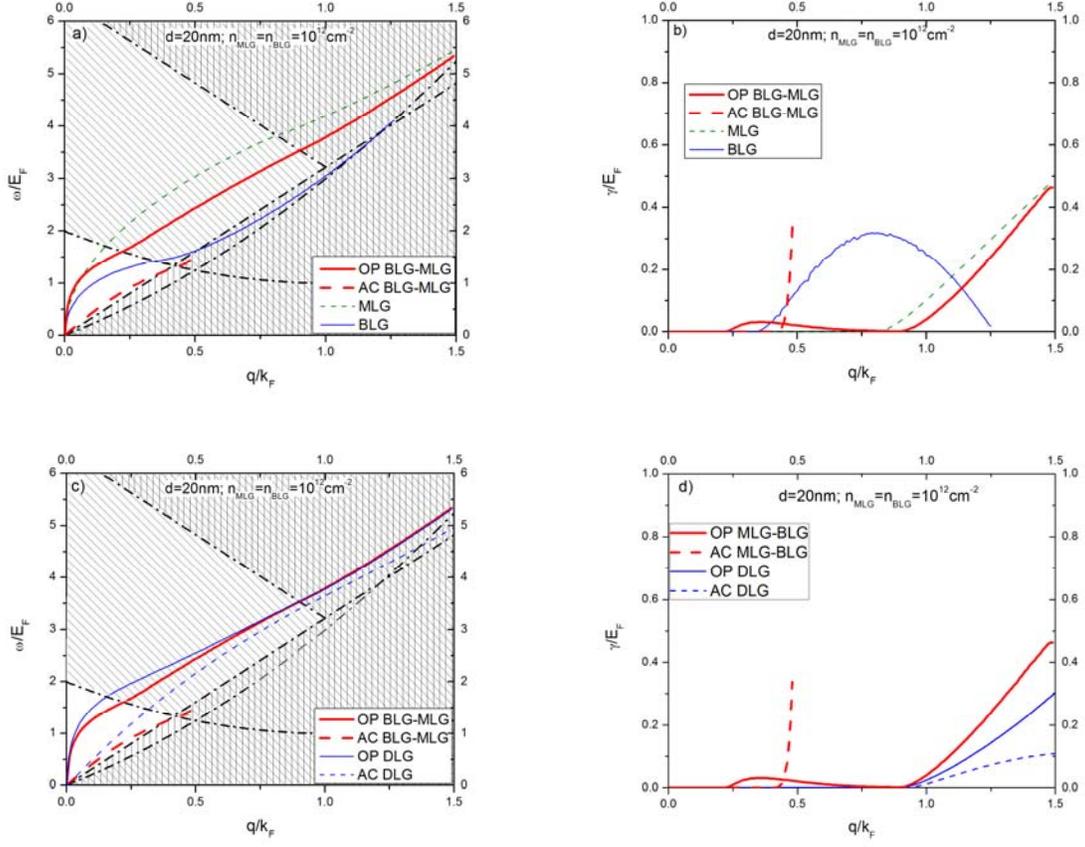

Fig. 3. Comparison of plasmon dispersion and damping rate in BLG-MLG, MLG and BLG systems ((a) and (b)) and in BLG-MLG and DLG double layers ((c) and (d)) for $\kappa_2 = \kappa_{Al_2O_3} = 9.1$, $d = 20$ nm and $n_{BLG} = n_{MLG} = 10^{12} \text{cm}^{-2}$. Average background dielectric constant used for BLG and DLG is $\bar{\kappa} = \frac{1}{2}(\kappa_1 + \kappa_3) = 2.4$. The dashed-dotted lines show the SPE boundaries of BLG and MLG (color online).

The effect of the interlayer distance is illustrated in Fig. 4 for $\kappa_2 = \kappa_{Al_2O_3} = 9.1$ using $d = 20$ nm and $d = 30$ nm. Fig. 4(a) shows that optical (acoustic) plasmon frequencies increase (decreases) with increasing interlayer distance. This behavior is similar to that of DLG showed in [15]. Fig. 4(b) indicates that the damping rate of acoustic mode is almost independent of the interlayer distance $d$ while that of optical mode decreases remakably as $d$ increases for wave-vectors inside interband continuum of BLG but out of the MLG SPE region.

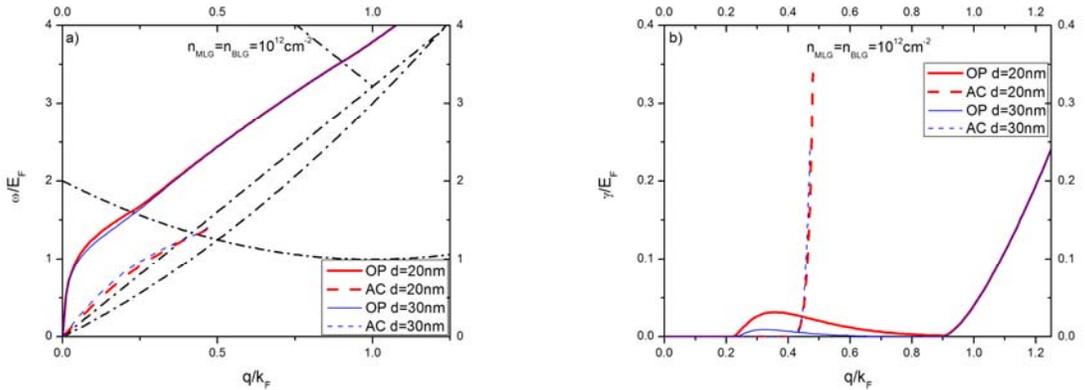

Fig 4. (a) Frequency and (b) damping rate of optical (solid curves) and acoustic (dashed curves) plasmon modes in a BLG-MLG system with $\kappa_2 = \kappa_{Al_2O_3} = 9.1$, $n_{MLG} = n_{BLG} = 10^{12} cm^{-2}$ for $d = 20nm$ and $d = 30nm$. The dashed-dotted lines show the SPE boundaries of BLG and MLG (color online).

To see the effect of different electron densities, we show in Fig. 5 plasmon dispersions and damping rate in a BLG-MLG system with $\kappa_2 = \kappa_{Al_2O_3} = 9.1$, $d = 20nm$ and $n_{BLG} = 10^{12} cm^{-2}$ for two cases $n_{MLG} = 0.5 n_{BLG}$ and $2 n_{BLG}$. It can be seen from the figure that frequencies and decay rates of optical (acoustic) plasmon modes depend strongly (slightly) on density ratio $n_{MLG} / n_{BLG}$ especially for large wave-vectors. The energy of both plasmon branches increases as carrier density in MLG increases.

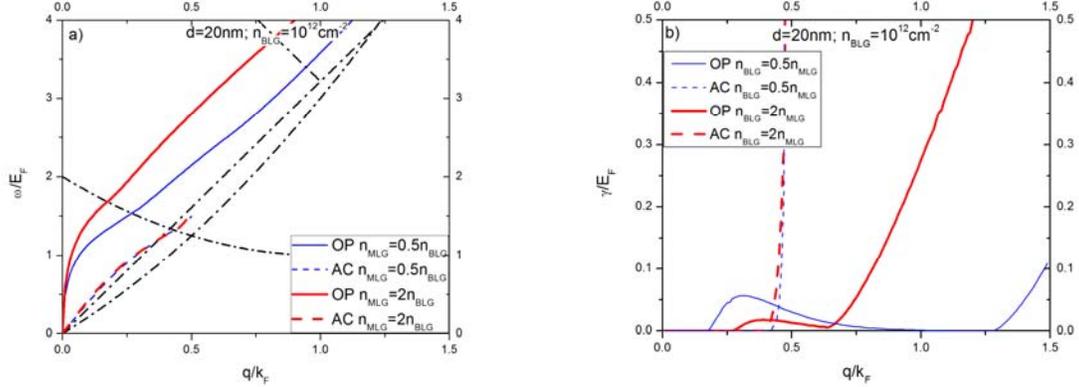

Fig 5. (a) Frequency and (b) damping rate of optical (solid curves) and acoustic (dashed curves) plasmon modes in a BLG-MLG system with $\kappa_2 = \kappa_{Al_2O_3} = 9.1$, $d = 20nm$ and $n_{BLG} = 10^{12} cm^{-2}$ (color online).

We now study the effect of nonhomogenous dielectric background by plotting in Fig. 6 (a) plasmon dispersions and (b) decay rates calculated for nonhomogenous dielectric background with $\kappa_2 = \kappa_{Al_2O_3} = 9.1$ (solid curves) and for homogenous dielectric background with an average permittivity $\bar{\kappa} = \frac{1}{2}(\kappa_1 + \kappa_3) = 2.4$ (dashed curves). Fig. 6(a) indicates that the inhomogeneity of the dielectric background decreases both optical and acoustic plasmon frequency. The effect of inhomogeneity of the background dielectric environment on properties of optical plasmon modes is more pronounced at larger wave-vectors.

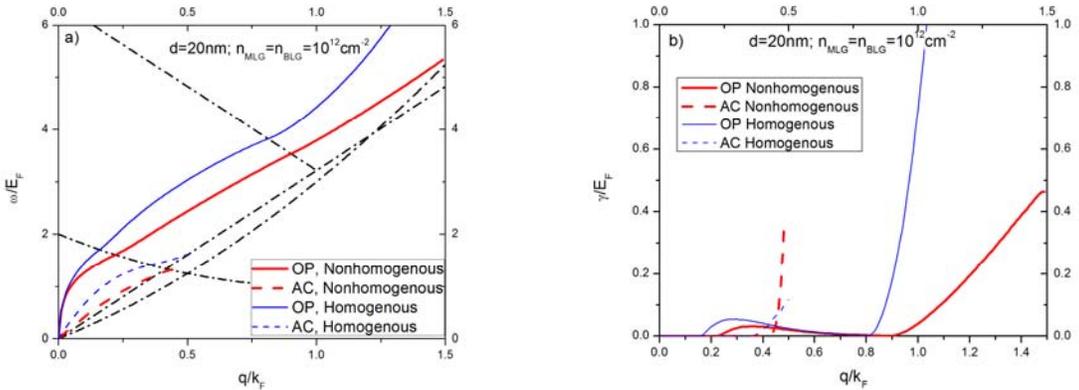

Fig. 6. Optical (upper curves) and acoustic (lower curves) plasmon modes calculated for nonhomogenous dielectric background with $\kappa_2 = \kappa_{Al_2O_3} = 9.1$ (solid curves) and for homogenous dielectric background with an average permittivity $\bar{\kappa} = (\kappa_1 + \kappa_3)/2 = 2.4$ (dashed curves) (color online).

The importance of the spacer on plasmon dispersions and damping rate are illustrated in Fig. 7 for $\kappa_2 = \kappa_{Al_2O_3} = 9.1$ (thick curves) and $\kappa_2 = \kappa_{hBN} = 3.0$ (thin curves). Optical (solid curves) and acoustic (dashed curves) plasmon modes are calculated for $n_{BLG} = n_{MLG} = 10^{12}\,cm^{-2}$ and $d = 20\,nm$. We see that the energy of both optical and acoustic branches decreases as the dielectric constant of spacer increases. At small $q$, the frequency of optical mode is almost independent of the spacer dielectric constant as shown by our analytical result given in Eq. (12).

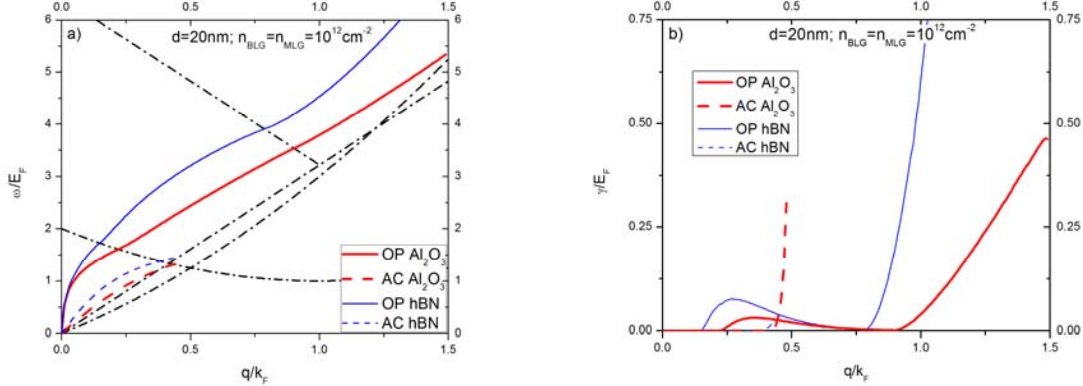

Fig. 7. Optical (solid curves) and acoustic (dashed curves) plasmon modes calculated for $Al_2O_3$ (thick curves) and hBN (thin curves) spacer with $n_{BLG} = n_{MLG} = 10^{12}\,cm^{-2}$ for $d = 20\,nm$ (color online).

## 4. Conclusion

In conclusion, the frequency and damping rate of plasmon oscillations in a BLG-MLG double layer at zero temperature are calculated for the first time using the RPA dielectric function and taking into account the nonhomogeneity of the dielectric background of the system. The analytical expressions for the plasmon frequency of optical and acoustic modes are derived using long wavelength expansions of response functions of MLG and BLG. We find that the frequency of optical (acoustic) plasmon branches decreases (increases) with increasing interlayer distance as in the case of DLG. It is shown that the energy of both plasmon branches decreases (increases) when the dielectric constant of spacer (density ratio $n_{MLG}/n_{BLG}$) increases. The frequency of optical mode is almost independent of the spacer dielectric constant at small $q$. Decay rates of optical (acoustic) plasmon modes are found to be strongly (slightly) dependent of the interlayer distance, inhomogeneity of the background, density ratio $n_{MLG}/n_{BLG}$ and spacer dielectric constant, especially at large wave-vectors.

**Acknowledgement**

This research is funded by Vietnam National Foundation for Science and Technology Development (NAFOSTED) under Grant number 103.01-2017.23.